\documentclass[prl,floatfix,reprint]{revtex4}
\usepackage{graphicx,amssymb,amsmath}

\begin{document}

\title{The Multilayer Temporal Network of Public Transport in Great Britain}

\author{Riccardo Gallotti\footnote{Correspondence to: rgallotti@gmail.com}} 
\affiliation{Institut de Physique Th\'{e}orique, CEA, CNRS-URA 2306, F-91191, Gif-sur-Yvette, France}
\author{Marc Barthelemy}
\affiliation{Institut de Physique Th\'{e}orique, CEA, CNRS-URA 2306, F-91191, Gif-sur-Yvette, France}

\begin{abstract}

Despite the widespread availability of information concerning Public Transport from different sources, it is extremely hard to have a complete picture, in particular at a national scale. Here, we integrate timetable data obtained from the United Kingdom open-data program together with timetables of domestic flights, and obtain a comprehensive snapshot of the temporal characteristics of the whole UK public transport system for a week in October 2010. In order to focus on the multi-modal aspects of the system, we use a coarse graining procedure and define explicitly the coupling between different transport modes such as connections at airports, ferry docks, rail, metro, coach and bus stations. The resulting weighted, directed, temporal and multilayer network is provided in simple, commonly used formats, ensuring easy accessibility and the possibility of a straightforward use of old or specifically developed methods on this new and extensive dataset.

\end{abstract}

\maketitle

\section*{Background \& Summary}

Public Transport is a fundamental service, provided in every country at various scales, and which answers to a mobility demand of a large share of the population. The quality of public transport systems directly influences the citizens' quality of life, by reducing travel-times, promoting social fairness and by improving the air quality in metropolitan areas. Increased public transportation investments also lead to a significant economic growth~\cite{APTA}. The increasing complexity of these systems and the lack of proper tools for their analysis, render the  managing tasks of transportation agencies harder. This is particularly true for large multi-modal systems, as single agencies often manage different separated parts of the network and both data handling and optimization have to cross hard organizational boundaries. In addition, from the users' point of view, the enormous amount of information available is such that the navigation in cities and planning of individual trajectories require now services offered by major information technology companies. 

Recently, network studies provided new approaches for the analysis of spatial~\cite{Barthelemy:2011}, temporal~\cite{Holme:2012} and multilayer networks~\cite{Kivela:2014, Boccaletti:2014}. Public transport networks represent a paradigmatical example of all these three categories (see Fig.~1 for a visualization obtained with the tool \cite{muxviz}), and we can gain a new understanding of these systems with these new methods \cite{DeDomenico:2014,scireppaper}. A large number of studies have been made considering only a single transportation mode~\cite{Latora:2001, Angeloudis:2006, Lee:2008, Derrible:2010, Roth:2012, Legara:2014, Sen:2003, Seaton:2004, Kurant:2006, Kurant:2006b, Li:2007, Drobritz:2009, Sienkiewicz:2005, Xu:2007, Chen:2007, vonFerber:2009} and a few studies only focused on the multimodal aspects, characterizing the resilience and navigability~\cite{DeDomenico:2014} or the temporal synchronization~\cite{Coffey:2012} transport network at a urban level (see Fig.~2 for an illustration in London).

In order to perform an extensive and comprehensive analysis at both the urban and inter-urban levels, we integrate timetable information for land and water transport, obtained from a rich open-data source, the United Kingdom's National Public Transport Data Repository (NPTDR - Data Citation 1), together with the schedules of all non-stop domestic flights in the United Kingdom provided by Innovata LLC~\cite{innovata}. This integration gives us the complete knowledge of British Public Transport, all modes included. 

\begin{figure*}[h!]
\begin{center}
\includegraphics[width=.9\linewidth]{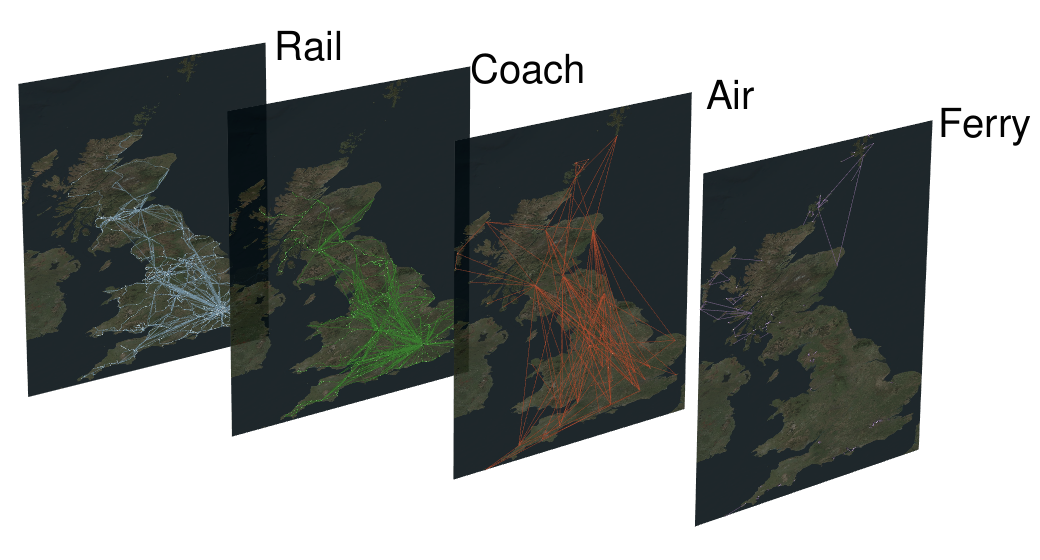}
\end{center}
\caption{National Public Transport Network in Great Britain. The dataset allows the study of the characteristics of the Public Transport Network at a national scale: the inter-urban connections are through Rail, Air and Coach layer. Air and Ferry connect mainland with the Northern Isles and the Hebrides. Metro and Bus are here not represented. (Rendered with MuxViz~\cite{muxviz})}
\end{figure*}

We use the multilayer network framework to identify the multi-modal features of this integrated transport system, and associate to each transportation mode a separate layer. We then aggregate stops associated to different modes into single network's nodes by using a hierarchical coarse-graining procedure that integrates and corrects the grouping information provided by the National Public Transport Access Nodes (NAPTAN) scheme~\cite{NaPTAN} used by the NPTDR. A given node then exists in different layers, and if it is used for different transport modes, the modal interconnection is represented by an inter-layer edge with a weight representing the walking time between the stops.

For all modes, the timetable information provides at each stop the departure and arrival times $t_{dep}$ and $t_{arr}$. As usually done in temporal networks \cite{Holme:2012}, we associate to each ride from a node $a$ to its consecutive neighbor $b$, a directed edge and instantaneous events $t=t_{dep}$ and $t=t_{arr}$ for departure and arrival of this ride. This edge is then considered as active during this trip and its weight is given by the travel-time $dt = t_{arr}-t_{dep}$. In addition, we also define a static (non-temporal) network, where the weight of each active edge is the minimal travel-time among all events that happened along it.

In order to ensure an easy access to the data, we write the static network information in Comma Separated Values files, using standard format of network theory: a file contains the list of nodes and another one contains the list of edges. The $\approx270K$ nodes distributed over 6 layers of the resulting network are geo-referenced and associated to the original NaPTAN metadata (see Table 1 for a summary). The $\approx470K$ intra-layer and inter-layer edges can be easily read from the multilayer edge-list. The temporal characteristics of the transport network are then listed in a separate file describing the $\approx130M$ events happening during a week of service in Great Britain during October 2010. Because of the dimension of the event list and the lack of established standard, we use a format optimizing the space requirements for this particular case.

\section*{Methods}

\begin{figure*}[h!]
\begin{center}
\includegraphics[width=.9\linewidth]{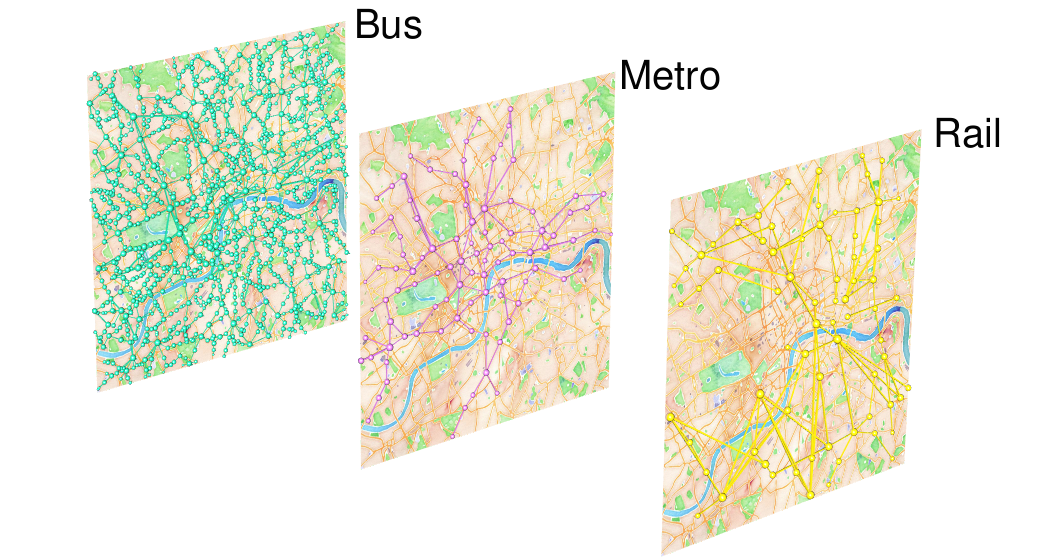}
\end{center}
\caption{Urban Public Transport Network in London. In general, the most widespread mode of transport at the urban scale is the Bus. Other layers like Ferry and Coach are sometimes present, but Rail and, when available, Metro represent the most viable alternatives to Buses for intra-urban trajectories~\cite{scireppaper}. (Rendered with MuxViz~\cite{muxviz})}
\end{figure*}

\subsection*{Original datasources}

Land and water Public Transport timetables are provided by the NPTDR (Data Citation 1) under Open Government licence. Snapshots of every journey are recorded for all services running in Great Britain (i.e. England, Scotland and Wales) during a full week in October for years ranging between 2004 and 2011. We choose the year 2010 as a working dataset as it was the most recent, fully consistent one (some more recent timetables are not updated for some areas in 2011). The raw files contain the information available in the travel-lines' web sites and call-centres during the selected week. For road transport, transportation agencies take into account the average traffic conditions at different hours and days in the design of timetables, and these data then implicitly contain the effects of congestion. The timetable information is provided within a resolution of the minute which we will therefore use in our dataset.

In the NPTDR data, the modes covered and identified are: bus, coach, (national) rail, ferry and metro (which includes underground, tram, light rail and non-national rail trains). All routes are originally referenced to stops defined using the NaPTAN scheme~\cite{NaPTAN}. In this scheme, every UK rail or metro station, coach terminal, airport, ferry terminal, bus stop or taxi rank is associated to at least one Stop Point. Each stop point is geo-referenced, has a detailed toponym and is identified by an ID denominated ATCOcode. For all bus stops, which represent the large majority of stops, the first three characters of the ATCOcodes can be linked to the Administrative Areas. In order to get a complete picture for all transportation modes, we use detailed schedules of all non-stop UK domestic flights, provided by Innovata LLC~\cite{innovata} for the week of 18-24 October 2010 (Innovata kindly accepted to grant us the rights of sharing these timetables and licensing this derived dataset as Open Data). Each of these flights has been associated to the Stop Points of the arrival and departure airport (and to a specific terminal whenever these were present). This was possible as each Stop Point contains in its ATCOcode the International Air Transport Association (IATA) code. Not all Stop Points that are defined are actually used. In our dataset, only those present in the timetables have been considered active and have therefore been taken into account. 

\subsection*{Timetables correction}
The original NPTDR data are not free from errors. For this reason, the timetables underwent a data-cleaning process:
\begin{itemize}
\item  A specific error was found in the Rail timetables. For some reasons, some unrecorded times were substituted with a '00000000'  string, which also represents a correct stop time at midnight. When an error happens at the arrival at a station B following station A, the passage time at B is estimated by adding to the passage time at A the average of the travel times recorded for all other correct A-B trips. This solution solved the problem almost entirely.
\item Further inconsistent stop times have been corrected by temporal interpolation whenever possible. This temporal interpolation assumes for instance that the travel-time (when known) $t_{AC}=t_{C}-t_{A}$ between A and C in a A-B-C trajectory has to be split in $t_{AC} = t_{AB}+t_{BC}$ where $t_{AB}/t_{BC} = d_{AB}/d_{BC}$, where $d_{IJ}$ is the euclidean distance between $I$ and $J$. More specifically, many inconsistencies were found in bus stop times: they have been considered wrong whenever two following stops are separated by a temporal interval of more than 2 hours (this also applies to the cases when a stop occurs at time which does not respect time causality). When the interpolation was not possible for a given event, we excluded it from the dataset. The fraction of events excluded under this rule is negligible when compared to the overall number of time events.
\end{itemize}

\subsection*{Coarse graining procedure}

Stop points are then organized in Stop Areas representing facilities (Airports, Bus/Metro/Coach/Railway Stations) or possible interchange points. Also Stop Areas are geo-referenced and identified by an ATCOcode (characterized by the character `G' at position four in the code). These Stop Areas have been taken as a basis for defining a multilayer network from the timetable data. Probably because of the bottom-up data entry process used for NaPTAN, the Stop Areas were not homogeneously defined. In particular, the spatial size can vary significantly and in some cases groups of stops can cover a distance of order 10 kilometres or more. In order to define consistently the nodes in the multilayer network where the inter-layer edges represent the inter-modal connections, we define inter-modal exchange nodes by correcting inconsistent NaPTAN Stop Areas using as a reference parameter a walking distance $wd$ = 500 m, used here as the maximal distance allowed to reach a bus stop by walking. We focused on bus stops as we identified them as the main source of errors, and we therefore removed bus stops from groups when they were out of the walking distance range and included them when they were within that range. Finally, the procedure aimed to maintain the hierarchy defined by the NaPTAN scheme: Air > Ferry > Rail > Metro > Coach > Bus and in the coarse graining we took into account this ranking where Airports are the most important locations and Bus stops the less important.

In more detail, this data-cleaning and aggregation procedure follows the steps listed hereafter:
\begin{itemize}
\item[i)] the Centres of an Area are identified by their latitude and longitude in the NaPTAN dataset. If not all Points are within a distance $wd$ from the Centre, the centre of mass of all Stop Points is computed as the point with latitude and longitude equal to the mean value of all Points latitudes and longitudes respectively. If all Points are then within a distance $wd$ from the centre of mass, it is used as the new Centre of the Area.
\item[ii)] Points where Air, Ferry, Rail, Metro and Coach stops occur are always kept in the Area, and Bus stop Points further than $wd$ from the Centre are removed;
\item[iii)] Areas containing only Bus stops Points are corrected by iteratively removing the farthest stop from the centre and recalculating the centre of mass, until they become contained in a circular area of radius $wd/2$ (thus implying a maximal distance between two points in this area less than $wd$);
\item[iv)] Airport Stop Points and Areas are joined together if they share the same IATA code;
\item[v)] All Airport Stop Points are `promoted' to Areas;
\item[vi)] The Heathrow Airport Stop Area is reconstructed with a specific rule (lines 777-790 of 1\_stops.py in Data Citation 2) as the Stop Area was incorrectly defined in the original dataset;
\item[vii)] Using the hierarchy mentioned above, all Areas include other Areas and non-bus stops Points of lower rank within a distance $wd$ from its Centre (the distance between Areas is defined as the distance between their centres);
\item[viii)] All remaining non-bus stops Points are `promoted' to Areas;
\item[ix)] Rail, Metro and Ferry Areas of same rank are merged if their distance is under $wd$ (Rail) or $wd/2$ (Ferry, Metro). Here the threshold $wd/2$ is chosen to avoid joining together consecutive Stops of Tube and Ferry lines in London;
\item[x)] All Areas can absorb a Bus stop Point if it is within a distance $wd/2$. In case of conflict, the Point is assigned to the closest Area;
\item[xi)] Areas containing only one Point are `declassed' to Points;
\item[xii)] A stop Point can absorb lower rank Areas/Points if it is within a distance $wd/2$ and thus becoming an Area (Coach and Bus Stops cannot absorb in this step);
\item[xiii)] We assign to each Area a representative Point, chosen at random between those with the higher rank;
\item[xiv)] We assign an areacode to each Area which corresponds to the first 3 digits of the ATCOcode of the majority of its bus stops. These 3 digits are associated to the Administrative Area in the NaPTAN scheme.
\end{itemize}

\subsection*{Defining the network}

The Areas defined by the procedure described above comprise nodes connecting the different layers of the multilayer network. The distance used for computing inter-layer weight is then computed as the average distance between all Points of the first layer and all Points of the second layer within these Areas. This walking time is obtained from this distance using a standard walking speed of $5$ km/h and converted in minutes. Time is described by an integer in our dataset and in order to avoid occasional divide-by-zero errors, we added 1 minute of minimal connection time to these connection times, which may reflect the time needed for getting off the vehicle, finding the vehicle of the following ride and getting on that second vehicle. We also took into account the 2 hours time requested for airport screening procedures, check-ins and luggage retrieval. These 2 hours of connection times are added to any connection towards the Air layer and 30 minutes are added to all connections from the Air layer.

Both the NAPTAN and the Innovata data use a format where each route is considered as active during certain days of the weeks. We prefer to define the temporal networks by translating directly this information into the existence or not of links between two nodes at a given moment measured in minutes from the 00:00 of Monday. This choice does not optimize memory space but allows for an extremely straightforward reading of the data. Not all studies need the temporal network information, and we therefore defined a weighted static network, where we associate to each edge a weight given by the minimal travel-time recorded during the considered week. We impose this minimal travel-time to be at least of one minute, again to avoid divide-by-zero errors for times defined with a 1 minute precision. We note that, because of the coarse-graining method used for defining nodes, self loop edges are present in our network as some routes could pass by more than one stop that have then been aggregated in the same node.

\subsection*{Code availability}

In order to ensure the replicability and reproducibility of our dataset, we share all codes that were developed for producing this dataset. The software is written in Python 2.7 and can be found in the repository indicated as Data Citation 2. The input files needed are the air timetables provided to us by Innovata LLC, which are included in our dataset (Data Citation 3) and the 2010 snapshot the NPTDR data (Data Citation 1). In particular: 
\begin{itemize}
\item{} the folder identified in the code as ``NPTDRTimetablesPath'' is contained in all the unzipped version of the files present in the folder October-2010/Timetable Data/CIF/National; 
\item{} the folder ``NaPTANPath''  contains the file present in the .zip file October-2010/NaPTANcsv.zip;
\item{} the folder ``INNOVATAtimetablesPath'' contains the file UKDOMESTICOCT10.csv that accompanies our dataset (Data Citation 3). 
\end{itemize}

The workflow consists in the following steps:
\begin{description}
\item[0\_correctRailTimetables.py] Correct the Rail timetables from the '0000000' error. It produces a corrected copy of the timetable in the original .CIF format, used in step 1.
\item[1\_stops.py]  Recognizes active stops, performs the stops' coarse-graining, associates nodes with areacode, corrects inconsistencies in all timetables, computes intra-layer distances. It produces a set of intermediate files (nodes list, events list, intra-layer edges list) used for the steps 2 and 3.
\item[2\_links.py] Sorts and rewrites the events list, compute the minimal traveltime for all edges. The output is a second version of the events list and a intra-layer edges list, used in step 3.
\item[3\_finalformat.py]  Computes the inter-layer traveltime, corrects the minimal traveltime when 0. The output is  the final format of the dataset.
\end{description}

The file layers.csv is simply typed in a text editor. Several parameters can be easily modified in this workflow. The walking distance $wd$ is defined in step 1. Walking speed, flight connection times, minimal connection time and the lower threshold to the minimal traveltime are defined in step 3. In addition, the same workflow can be also applied for all years where the NPTDR data are available (2004-2011).

\section*{Data Records}

\begin{table}[h]
\begin{center}
\begin{tabular}{| l | r | c |r |}
\hline
File & Lines & Header & Size \\
\hline
layers.csv 	& 7			& Yes	&  76B \\
nodes.csv 	& 267.032		& Yes	& 12MB \\
edges.csv 	& 475.503		& Yes	& 14MB \\
events.txt		& 465.216		& No		& 893MB \\	
\hline 
\end{tabular}
\end{center}
\caption{Dataset dimensions.}
\end{table}

This dataset is stored as a single zip file at the Dryad Digital Repository (Data Citation 3). It describes the Public Transport Network of Great Britain by using a multilayer node-list and edge-list, where each layer is associated to a mode of transport. Each node is geo-referenced, thus defining a spatial network, and associated to a wide list of administrative meta-data. 
For each edge, we indicate the minimal travel-time in minutes. In the case of inter-layer edges, their weight is defined as the walking time needed for the inter-modal connection. For each intra-layer edge, we also specify a list of temporal events representing the vehicles' rides along the edges, with origin time and travel duration expressed in minutes with a precision of one minute. 
In multilayer networks, each node may have several `copies' in different layers. For that reason, nodes are identified by two numbers, one specifying the node itself, and another the layer where it belongs. As a consequence, the edges are identified by four numbers, two for the origin and two for the destination (node and layer). The ordering of the fields in edge-list has been chosen to conform to the tensorial notation widely used for this type of networks.
There is no established standard format for the temporal event list, and we therefore decided to use a format, derived from the idea of adjacency lists, which has been specifically adapted to this dataset.


All files are plain ASCII text files. Each row represents an item and for all files (except the file events.txt) the number of fields and their order is the same. Conversely, a non-standard format has been used for encoding the events information (see below).
\\ \\
\noindent
\textbf{layers.csv} In this network, each layer represents a mode of transport.
\begin{itemize}
\item{\em layer}: numerical id for each layer;
\item{\em layerLabel}: mode of transport associated to the layer.
\end{itemize}

\noindent
\textbf{nodes.csv} The same node, identified by the same {\em nodeid}, may exist in different layers and is represented once in each layer where it appears.
\begin{itemize}
\item{\em nodeid}: numerical id for each node;
\item{\em layerid}: layer when the node appears;
\item{\em lat}: latitude;
\item{\em lon}: longitude;
\item{\em areacode}: using this code one may link node to the administrative area information contained in `Admin Areas.csv' (where this key is called {\em ATCOcode}) and then to `Travel Region.csv' through the {\em traveline Region ID} key;
\item{\em atcocode}: the ATCOcode of a stop Area if its 4th character is `G', or of a stop Point if is `0'. With this code one may link nodes to the metadata of `Stops.csv' for Points and `Groups.csv' for Areas.
\end{itemize}

\noindent
\textbf{edges.csv} Edges are directed and weighted. They can be either intra-layer, between different nodes in the same layer, or inter-layer, between the same node in different layers. We associate a weight to each edge which is given by the minimal travel time from the origin to the destination, in minutes. Edges are listed following the layers' hierarchy.

\begin{itemize}
\item{\em ori\_node}: origin node;
\item{\em des\_node}: destination node;
\item{\em ori\_layer}: origin layer;
\item{\em des\_layer}: destination layer;
\item{\em minutes}: minimal travel time between origin and destination in minutes;
\item{\em km}: the euclidean distance between origin and destination in kilometres.
\end{itemize}

\noindent
\textbf{events.txt} Edges are listed following the layers' hierarchy. The bus events are by far the most numerous (more than $90\%$ of the total), and the reading can be interrupted at the first edge belonging to this layer whenever the bus information is not needed. The format for this file is the following:\\

\noindent
{\em ori\_node, des\_node, ori\_layer, des\_layer, $t_1$, $dt_1$, $t_2$, $dt_2$, \dots, $t_n$, $dt_n$ }\\

\noindent
It would have been extremely space demanding to write this file as a simple event list, and for this reason this file is not written as a standard .csv file: we still use commas as delimiters for each line of the text, but the number of columns is different in each row. We decided to use the same structure of adjacency lists and list the events as a succession in the same row.  Each row represents an edge, described in the first 4 fields. The subsequent fields are a list of events. Each edge has a different number of events, and therefore each line a different number of fields. Each event i is identified by two values, $t_i$ and $dt_i$, and represents a ride starting at time $t_i$ and of duration $dt_i$. Both times are in minutes. $t_i$ is defined as minutes starting from the 00:00 of Monday.

\subsection*{NaPTAN and NPTG metadata}

We include the original files from the National Public Transport Access Node (NaPTAN) and National Public Transport Gazeetteer (NPTG). We encourage the reader to refer to the comprehensive guide ``naptanschemaguide-2.5-v0.67.pdf'' available online~\cite{naptanguide} that can be found on the NaPTAN website (http://www.naptan.org.uk/schema/schemas.htm).
Here the list of fields for each of these files:
\\

\noindent
\textbf{Stops.csv}

\noindent
ATCOCode,
GridType,
Easting,
Northing,
Lon,
Lat,
CommonName,
Identifier,
Direction,
Street,
Landmark,
NatGazID,
NatGazLocality,
ParentLocality,
GrandParentLocality,
Town,
Suburb,
StopType,
BusStopType,
BusRegistrationStatus,
RecordStatus,
Notes,
LocalityCentre,
SMSNumber,
LastChanged.
\\

\noindent
\textbf{Groups.csv}

\noindent
GroupID,
GroupName,
Type,
GridType,
Easting,
Northing,
Lon,
Lat,
LastChanged.
\\

\noindent
\textbf{Admin Areas.csv}

\noindent
Admin Area ID,
Admin Area Name,
Traveline Region ID,
Country,
ATCO Code,
Call Centre ID,
Date of Issue,
Issue Version.
\\

\noindent
\textbf{Travel Region.csv}

\noindent
Traveline Region ID,
Region Name,
Primary URL,
Secondary URL,
Tertiary URL,
Date of Issue,
Issue Version,
JW Version.

\subsection*{Innovata LLC timetables}

To ensure the reproducibility of our dataset, we include the original timetable as provided by Innovata LLC~\cite{innovata}.
\\

\noindent
\textbf{UKDOMESTICOCT10.csv}

\noindent
Mktg Al,
Op Al,
Orig,
Dest,
Kilometers,
Flight,
Stops,
Equip,
Seats,
Dep Term,
Arr Term,
Dep Time,
Arr Time,
Block Mins
Arr Flag,
Op Days,
Ops/Week,
Seats/Week.

\section*{Technical Validation}

The reliability of the spatial and temporal information largely depends on the reliability of the source data provided by the transport agencies to the NPTDR. Further limits come from the multilayer aspects of the network we have defined. The absence of other data sources to compare our results with, limit our possibilities for validating the technical quality of our dataset. For this reason, we propose in the Supplementary Information File a statistical characterization of all quantities represented in the dataset. We use such analysis as `sanity checks' to support the correctness of the information provided. In addition, we study the characteristics of time-respecting shortest paths through London, defined as the quickest journeys along a a sequence of connections with non-decreasing times~\cite{Holme:2012} and studied by us in a recently published paper~\cite{scireppaper}. The apparent correctness of the  large majority of these paths provide further support about the reliability of the temporal events and about our definition of the multilayer network.

\subsection*{Spatial aspects}

We verify the consistency of the coarse-graining procedure, nodes position Stop Area centres and Stop Points by visualizing them on a satellite map. The positioning was always found reasonable at that scale ($\approx$ 500 m). For this reasons, we are rather confident that in the original files the latitude, longitude, easting and northing fields are correct. From these coordinates, we measured the intra-layer distances, whose values appear to be sensible, and the shape of the probability distribution also correctly matches the walking distance constraint we have introduced (seeSupplementary Information Supplementary Information). In the Supplementary Information, one can also observe that the final node positions reasonably cover the shape of Great Britain, and that no transcription errors appear to have been made. From these coordinates, we compute the edges lengths, whose values falls in reasonable ranges for each layer (note that a set of relatively short ($<$100 kms) edges are present in the air layer: they are mostly associated to low capacity flights or travels among the Orkney Islands).

\subsection*{Temporal aspects}

In our procedure, we correct some errors (manifestly wrong or impossible times), directly enhancing the quality of the data the original source. Nevertheless, from the statistical analysis we perform in the Supplementary Information, it emerges the most relevant limit of the dataset: the low reliability of the short traveltimes. Indeed, a fraction of 1\% of the bus layer's edges have a max travel speed (edge length/minimal traveltime) higher than 90 km/h. The most extreme of those speeds are associated to extremely short traveltimes. This issue is inherited from the original NPTDR database and is a consequence of the low precision of bus timetables. This problem is reflected on the events data and magnified if one consider the max travel speed (thus our choice of a lower bound of 1 minute for the minimal traveltime, to avoid divisions by zero).
 
Although traveltimes on the single edges are not totally reliable, these errors become less relevant when computing total time along a trajectory. The longer the trajectory the more the error is negligible. In particular, we verified that, in London, the distribution of the speeds for short time-respecting paths is short tailed, and with a relatively reasonable extreme value ($\approx$60km/h), already for extremely short trajectories of covering distances in the interval [500,750] m. 

\subsection*{Multilayer aspects}

In the Supplementary Information, we check that the choice for the walking distance between different layers is reasonable. Naturally, small errors consequent to the coarse-graining procedure are unavoidable, but most of the more complex connection sites (airports, large stations) were already correctly defined in the NaPTAN scheme. This has been verified by visualizing the Stop Points on a satellite map. Clearly, taking walking times proportional to euclidean distances is only an approximation as it does not take into account the actual shape of the connection facilities. In order to check the overall quality of the connection definitions, we performed a visual comparison between time-respecting shortest paths through London. We took as a starting time Monday at 08:00, and compared the results given for the same origin and destination in the Google Maps website. Although the results always presented some deviations, probably because of the differences between data sources and choices made by Google for computing shortest path along the street and transport networks, trajectories were similar for large portions to those obtained from our data (with the option `less walking' taken).

\section*{Acknowledgments} 

The authors are supported by the European Commission FET-Proactive project PLEXMATH (Grant No. 317614). We thank Manlio De Domenico, Mikko Kivel\"a and Mason Porter for useful feedbacks on the format of this dataset.

\section*{Author Contributions}
RG designed the dataset, processed the data and wrote the paper. MB designed the dataset and wrote the paper.

\section*{Competing financial interests}
The authors declare no competing financial interests.

\section*{Data Citations}

1. Transport Direct.
\emph{data.gov.uk} http://data.gov.uk/dataset/nptdr (Date of access: 06/06/2014) (2010).

2. Gallotti, R. \& Barthelemy, M.
\emph{figshare.} doi:10.6084/m9.figshare.1249862 (2014).

3. Gallotti, R. \& Barthelemy, M.
\emph{Dryad Digital Repository.} doi:10.5061/dryad.pc8m3 (2014).

\section{Supplementary Information}

In this Supplementary Information, we provide a statistical analysis
of the quantities described in this dataset. The goals of this
analysis is to check the limits of the data reliability and at the
same time offering to future users a first insight into the
characteristics of the Public Transport Network.

\begin{figure}[h!]
\centerline{
\includegraphics[width=1.1\linewidth]{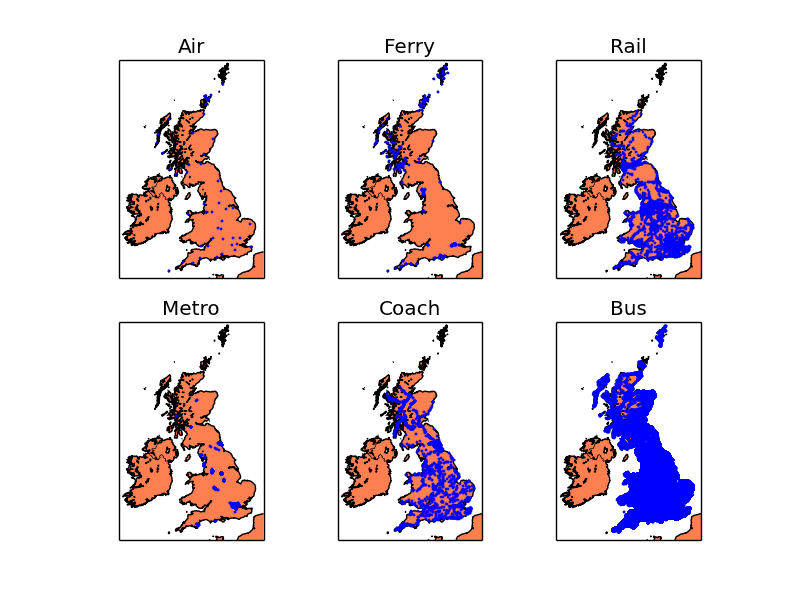}
}
\caption{{\bf Location of the nodes of each layer.} The distribution
  of points matches the map of Great Britain. The Bus layer is
  covering the largest fraction of the island, while Ferry and Metro
  appear to be more concentrated in specific areas: the islands in the
  north of Scotland for the Ferry layer and the large urban areas of
  England for the Metro. The Coach, Rail and Air are more
  homogeneously spread, as they serve the purposes of inter-city
  connection across the whole country. (Figure produced with the
  Basemap python library).}
\label{nodes}
\end{figure}

\begin{figure*}[h!]
\centerline{
\includegraphics[width=.9\linewidth]{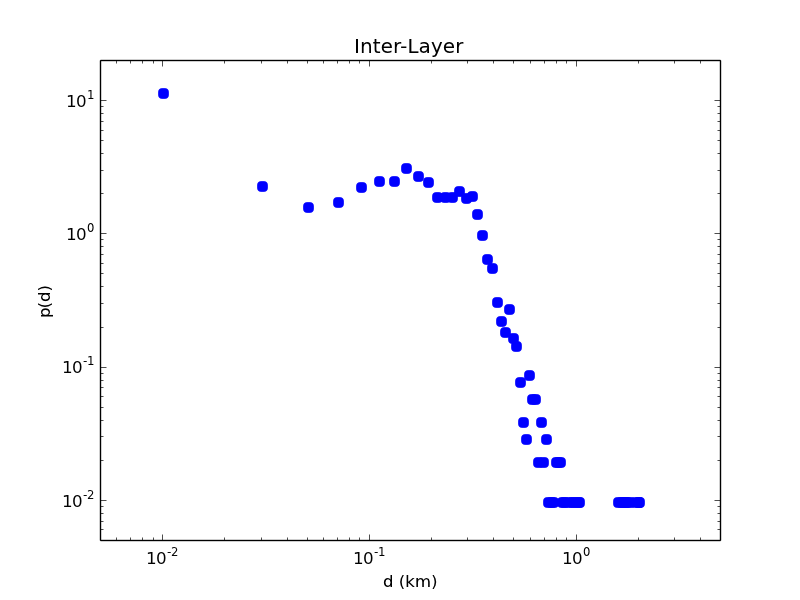}
}
\caption{{\bf Inter-layer distance distribution.} The distribution
  is relatively uniform for values under 300 meters, and then quickly
  decays, having a maximum value of $\approx$2 km. These values are
  in agreement with the imposed maximal walking distance of 500 m imposed
  to low rank facilities and to the typical size of large facilities
  such as airports (for example, in order to reconstruct Heathrow
  airport we
  set a maximal distance of 3 km).  We do not study the Inter-Layer
  travel time as it follows directly from the distance, the walking
  speed, and the minimal connection time.}
\label{interD}
\end{figure*}

\begin{figure*}[h!]
\centerline{
\includegraphics[width=1.2\linewidth]{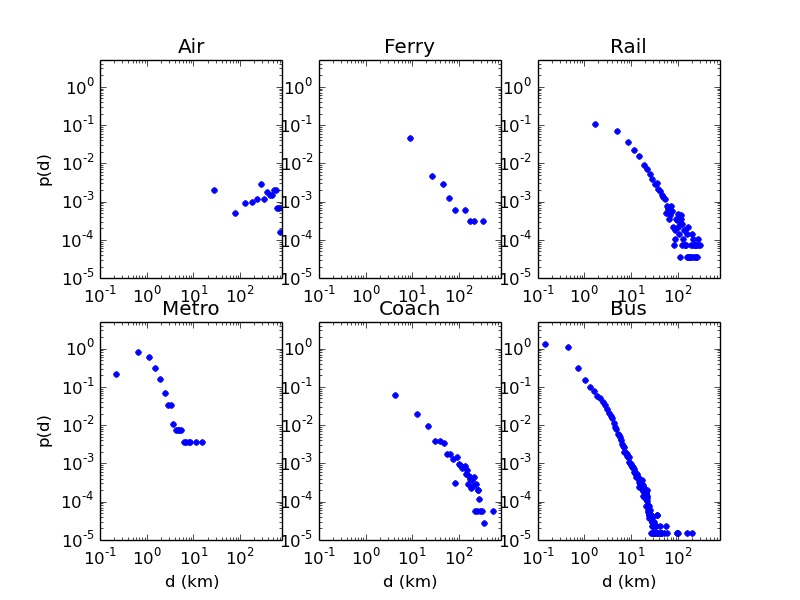}
}
\caption{{\bf Intra-layer distance distribution.} Different layers
  have different characteristic distances. The largest part of Bus and
  Metro edges are short ($<$2 km), Rail edges appear above a minimal distance of $\approx$2 km, followed by Coach
  ($\approx$5 km) Ferry ($\approx$10 km). There is a significant
  fraction of Air edges under 100 km, mostly associated with low
  capacity or local flights between the Orkney
  Islands. Concerning the upper bound of the distribution, the longest
  distances are covered by flights. The Ferry, Rail and Coach layer
  appear all limited to maximal travels of $\approx$300 km. Metro
  edges are limited to the size of urban areas ($\approx$10 km),
  while Bus can observed up to distances of order  $\approx$50 km.}
\label{intraD}
\end{figure*}

\begin{figure*}[h!]
\centerline{
\includegraphics[width=1.2\linewidth]{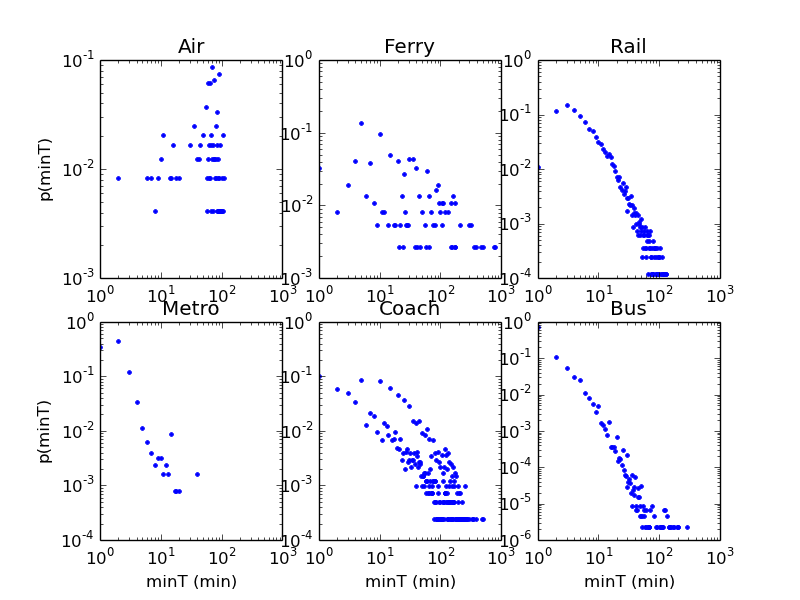}
}
\caption{{\bf Intra-layer minimal traveltimes distribution.} Comparing
  traveltimes is less straightforward than comparing distances, as
  each layer has different typical speeds. In the Air, Rail and Bus
  the maximum traveltimes appear to be limited to $\approx$ 2h, while
  for Metro the value is much lower ($\approx$20 min) and for Ferry
  and Coach higher.  We observe a large number of minimal traveltimes
  of 1 minute in the Bus layer. The peculiar flight with a minimal
  time of 2 minutes  has been identified as a small 9 seats aircraft covering a
  trajectory of 13 kms. The sparsity of the distributions, particularly
  noticeable for Air, Coach and Ferry is due to the preference of
  timetable designer for round number as, for instance, multiples of
  5.}
\label{minT}
\end{figure*}

\begin{figure*}[h!]
\centerline{
\includegraphics[width=1.2\linewidth]{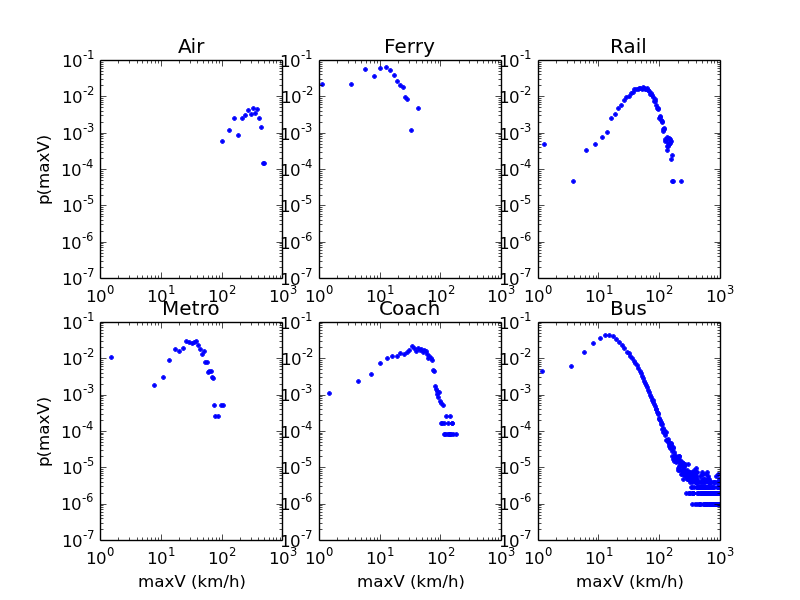}
}
\caption{{\bf Distribution of the maximum speeds.} All the distributions
  present a maximum value consistent with what could be expected for
  the different transportation modes. Air: $\approx$ 350 km/h. Ferry:
  $\approx$ 15 km/h. Rail: $\approx$ 60 km/h. Metro: $\approx$ 30
  km/h. Coach: $\approx$ 40 km/h. Bus: $\approx$ 15 km/h. We observe
  some cases impossible speeds for the Metro, Coach and Bus layers
  (with particularly extreme values in the latter: 7\% of the
  bus edges have a maximum speed higher than 50km/h, 1\% over
  90km/h). As we see in figure \ref{tSpeed}, these extreme values are
  associated with errors in very short travel times.}
\label{maxS}
\end{figure*}

\begin{figure*}[h!]
\centerline{
\includegraphics[width=1.2\linewidth]{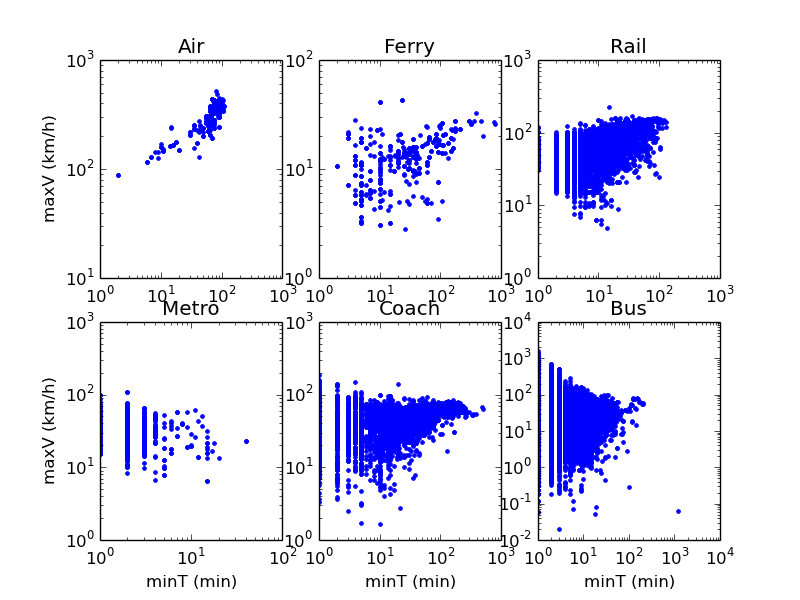}
}
\caption{{\bf Maximum speed versus minimal traveltime.}  As expected, we
  observe that for longer travel times, the speed increases. This behaviour is
  probably hidden in the sparsity of points for short times in the
  cases of the Metro, Coach and Bus layers. We see here that almost all high
  speed edges pointed out in figure \ref{maxS} are associated to short
  traveltimes.}
\label{tSpeed}
\end{figure*}

\begin{figure*}[h!]
\centerline{
\includegraphics[width=1.2\linewidth]{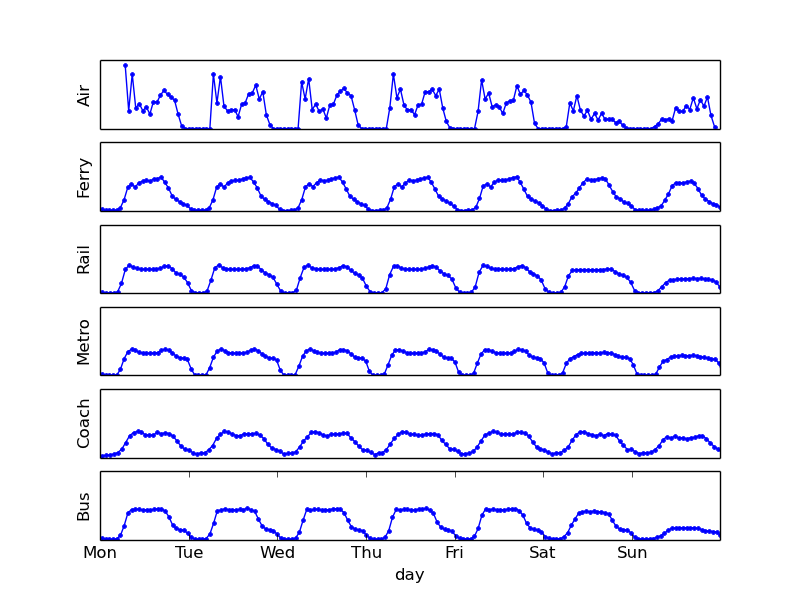}
}
\caption{{\bf Distribution of the events times in the week.} As we see,
  events are spread during the days consistently with a
  periodical daily urban activity cycle. Weekdays are mostly similar
  to each other, while peculiar changes appear in schedules during the weekend.}
\label{eventTime}
\end{figure*}

\begin{figure*}[h!]
\centerline{
\includegraphics[width=1.2\linewidth]{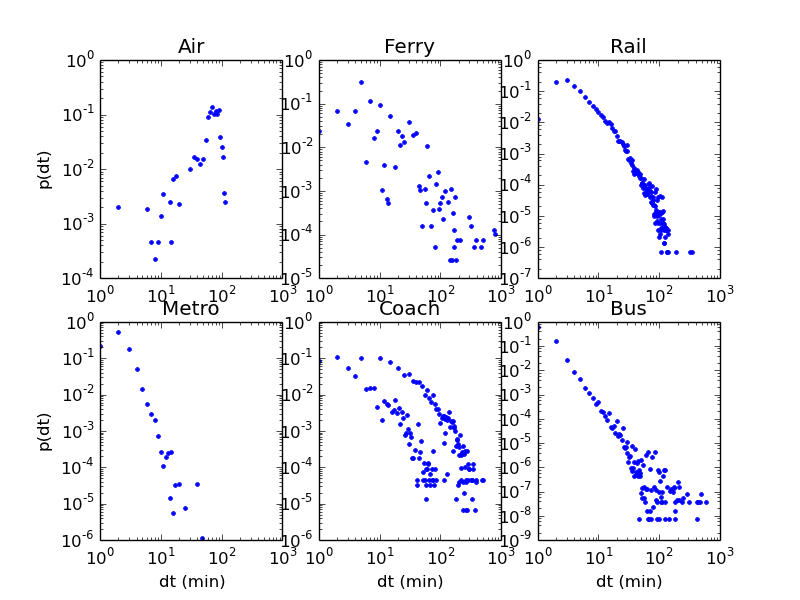}
}
\caption{{\bf Distribution of the events' travel times.} Events'
  travel times present the same problems as minimal
  travel times. In some cases, events share the same time leading to
  too large or even impossible velocities. Indeed, in this
  case, travel times equal to zero have been kept in order to maintain the time
  causality of the sequences of events, implying that infinite speeds are a
  priori possible on single edges.}
\label{eventTraveltime}
\end{figure*}

\end{document}